\documentclass[twocolumn,showpacs,preprintnumbers,amsmath,amssymb]{revtex4}

\usepackage{graphicx}
\usepackage{dcolumn}
\usepackage{bm}
\usepackage{amssymb}
\usepackage{amsfonts}
\usepackage{amsmath}

\usepackage{color}

\definecolor{nv}{rgb}{0.1,0.1,0.6}
\definecolor{pr}{rgb}{0.2,0.1,0.5}
\definecolor{mg}{rgb}{0.4,0.0,0.4}

\newcommand{\beq}{\begin{equation}}
\newcommand{\eeq}{\end{equation}}
\newcommand{\beqy}{\begin{eqnarray}}
\newcommand{\eeqy}{\end{eqnarray}}
\newcommand{\beqyn}{\begin{eqnarray*}}
\newcommand{\eeqyn}{\end{eqnarray*}}
\newcommand{\nl}{\newline}

\newcommand{\bs}{\begin{slide}}
\newcommand{\es}{\end{slide}}
\newcommand{\bc}{\begin{center}}
\newcommand{\ec}{\end{center}}
\newcommand{\bmin}{\begin{minipage}}
\newcommand{\emin}{\end{minipage}}

\newcommand{\ud}{\mathrm{d}}

\newcommand{\bi}{\begin{itemize}}
\newcommand{\ei}{\end{itemize}}

\begin{document}

\preprint{APS/123-QED}

\title{Comment on ``Proton Spin Structure from Measurable Parton Distributions" by Ji, Xiong and Yuan (PRL109, 152005 (2012))}

\author{Elliot Leader}
\affiliation{
Imperial College London\\ Prince Consort Road, London SW7 2AZ }

\author{C\'edric Lorc\'e}
\affiliation{IPNO, Universit\'e Paris-Sud, CNRS/IN2P3, 91406 Orsay, France\\
        and LPT, Universit\'e Paris-Sud, CNRS, 91406 Orsay, France}


\date{\today}

\begin{abstract}
We show that the recent claim that the expression  $\frac{1}{2 }\int_{-1}^{1} dx x \left[H_q(x,0,0)+E_q(x,0,0)\right]$,   involving the generalized parton distributions $H$ and $E$, measures the transverse angular momentum of quarks in a transversely polarized nucleon, is incorrect.
\end{abstract}

\maketitle

Some time ago Ji \cite{Ji:1997pf, Ji:1996ek, Ji:1996nm}, using the \emph{Belinfante version} of the angular momentum operator, derived a beautiful relation between the quark angular momentum and Generalized Parton Distributions (GPDs). In these papers the relation was written as
\beq \label{Ji}  J_q  =\frac{1}{2 }\int_{-1}^{1} dx x \left[H_q(x,0,0)+E_q(x,0,0)\right], \eeq
and for a decade and a half the quantity $J_q$ has been almost universally interpreted as the expectation value of the longitudinal component of the quark angular momentum in a longitudinally polarized nucleon, \emph{i.e.} for a nucleon moving along the $z$-direction
\beq \label{JiL} J_q = \langle \langle \,\, J^z_q\, \,\rangle  \rangle_L . \eeq

Inspired by the impact-parameter explanation of $J_q$ proposed by Burkardt \cite{Burkardt:2005hp}, Ji, Xiong and Yuan \cite{Ji:2012sj} show that a partonic interpretation of the RHS of Eq.~(\ref{Ji}) can be obtained, and state that $J_q$ measures the expectation value of the transverse angular momentum of the quarks in a nucleon polarized in the transverse direction $i$. What they claim to prove  is that
\beq \label{JXY} J_q \propto \langle \langle \,\, J^{+i}_q\, \,\rangle  \rangle_{T_i}  \eeq
where
\beq \label{J} J^{+i}_q=\int\ud x^-\ud^2x_\bot\,M^{++i}_q(x) \eeq
with, in terms of the Belinfante energy-momentum tensor density,
\beq \label{EMT} M^{\mu\rho\sigma}_q(x)= x^\rho \,T^{\mu\sigma}_q(x) - x^\sigma \,T^{\mu\rho}_q(x). \eeq

Since $J^{+i}$ is a leading-twist operator, it is clear that such a simple partonic interpretation should exist. However, Ji, Xiong and Yuan misleadingly interpret $J^{+i}$ as the transverse \emph{angular momentum} operator. Indeed, in light-front quantization the role of time is taken by $x^+$, so that $J^{+i}$ is the light-front transverse \emph{boost} operator. In terms of the more conventional instant-form boost ($K^i$) and rotation ($J^i$) operators (see \emph{e.g.} Refs.~\cite{Kogut:1969xa,Weinberg:1995mt,Brodsky:1997de}), the light-front transverse boost operators read
\beq \label{LFB} J^{+1}= \frac{1}{\sqrt{2}} (K^1+J^2), \qquad J^{+2}= \frac{1}{\sqrt{2}} (K^2-J^1),\eeq
while the light-front transverse angular momentum operators are given by
\beq \label{LFS} J^{-1}= \frac{1}{\sqrt{2}} (K^1-J^2), \qquad J^{-2}= \frac{1}{\sqrt{2}} (K^2+J^1).\eeq
The light-front transverse boosts ($J^{+i}$) are kinematic operators and therefore leading-twist, while the light-front transverse angular momenta ($J^{-i}$) are dynamical operators and therefore higher-twist. It is also easy to see that the quark and gluon spin operators in the $A^+=0$ gauge \cite{Jaffe:1989jz}
\beq\begin{aligned} M^{\mu\rho\sigma}_{q,\text{spin}}&=\frac{1}{2}\,\epsilon^{\mu\rho\sigma\nu}\,\overline\psi\gamma_\nu\gamma_5\psi,\\
M^{\mu\rho\sigma}_{g,\text{spin}}&=-2\,\text{Tr}\left[F^{\mu\rho}A^\sigma-F^{\mu\sigma}A^\rho\right]
\end{aligned}\eeq
contribute to $J^{-i}$ and not to $J ^{+i}$.

Any genuine transverse angular momentum sum rule is expected to have a frame dependence. The reason is simply the well-known fact that boosts and rotations do not commute. One consequence of this is that special relativity naturally induces spin-orbit correlations. Obviously, there cannot be any spin-orbit correlation with the longitudinal polarization, which is the reason why the longitudinal angular momentum sum rule is frame independent. On the contrary, the transverse polarization is correlated with the momentum, which is at the origin of the frame dependence of the transverse angular momentum sum rule.

In conclusion the Ji, Xiong and Yuan partonic interpretation has nothing to do with angular momentum.
One cannot  simply interpret $\frac{x}{2}\left[H_q(x,0,0)+E_q(x,0,0)\right]$ as the density of quark transverse angular momentum in a transversely polarized nucleon. A genuine transverse angular momentum sum rule naturally involves frame dependence, owing to the fact that boosts and rotations do not commute. Since the transverse angular momentum is a dynamical operator in light-front quantization, no simple partonic interpretation is expected. On the contrary, a simple partonic interpretation does exist for the longitudinal component of angular momentum in terms of Wigner distributions \cite{Lorce:2011kd, Lorce:2011ni, Lorce:2012rr, Lorce:2012ce, Hatta:2011ku, Ji:2012vj, Ji:2012ba}, precisely because it is a kinematic operator.\nl
 Finally,  one of us (E.L.) \cite{Leader:2011cr} recently derived a relation for the instant-form transverse component of the quark angular momentum  in a transversely polarized nucleon in terms of the GPDs $H$ and $E$, and in passing, checked that Eq.~\eqref{Ji} is indeed correct for the longitudinal case, with the identification $J_q = \langle \langle \,\, J^z_q\, \,\rangle  \rangle_L $. \nl

 E. L. thanks Ben Bakker for comments about light-front quantization.

\bibliography{Elliot_General}

\begin{thebibliography}{17}
\expandafter\ifx\csname natexlab\endcsname\relax\def\natexlab#1{#1}\fi
\expandafter\ifx\csname bibnamefont\endcsname\relax
  \def\bibnamefont#1{#1}\fi
\expandafter\ifx\csname bibfnamefont\endcsname\relax
  \def\bibfnamefont#1{#1}\fi
\expandafter\ifx\csname citenamefont\endcsname\relax
  \def\citenamefont#1{#1}\fi
\expandafter\ifx\csname url\endcsname\relax
  \def\url#1{\texttt{#1}}\fi
\expandafter\ifx\csname urlprefix\endcsname\relax\def\urlprefix{URL }\fi
\providecommand{\bibinfo}[2]{#2}
\providecommand{\eprint}[2][]{\url{#2}}

\bibitem[{\citenamefont{Ji}(1998)}]{Ji:1997pf}
\bibinfo{author}{\bibfnamefont{X.-D.} \bibnamefont{Ji}},
  \bibinfo{journal}{Phys. Rev.} \textbf{\bibinfo{volume}{D58}},
  \bibinfo{pages}{056003} (\bibinfo{year}{1998}), \eprint{hep-ph/9710290}.

\bibitem[{\citenamefont{Ji}(1997{\natexlab{a}})}]{Ji:1996ek}
\bibinfo{author}{\bibfnamefont{X.-D.} \bibnamefont{Ji}},
  \bibinfo{journal}{Phys. Rev. Lett.} \textbf{\bibinfo{volume}{78}},
  \bibinfo{pages}{610} (\bibinfo{year}{1997}{\natexlab{a}}),
  \eprint{hep-ph/9603249}.

\bibitem[{\citenamefont{Ji}(1997{\natexlab{b}})}]{Ji:1996nm}
\bibinfo{author}{\bibfnamefont{X.-D.} \bibnamefont{Ji}},
  \bibinfo{journal}{Phys. Rev.} \textbf{\bibinfo{volume}{D55}},
  \bibinfo{pages}{7114} (\bibinfo{year}{1997}{\natexlab{b}}),
  \eprint{hep-ph/9609381}.

\bibitem[{\citenamefont{Burkardt}(2005)}]{Burkardt:2005hp}
\bibinfo{author}{\bibfnamefont{M.}~\bibnamefont{Burkardt}},
  \bibinfo{journal}{Phys.Rev.} \textbf{\bibinfo{volume}{D72}},
  \bibinfo{pages}{094020} (\bibinfo{year}{2005}), \eprint{hep-ph/0505189}.

\bibitem[{\citenamefont{Ji et~al.}(2012{\natexlab{a}})\citenamefont{Ji, Xiong,
  and Yuan}}]{Ji:2012sj}
\bibinfo{author}{\bibfnamefont{X.}~\bibnamefont{Ji}},
  \bibinfo{author}{\bibfnamefont{X.}~\bibnamefont{Xiong}}, \bibnamefont{and}
  \bibinfo{author}{\bibfnamefont{F.}~\bibnamefont{Yuan}},
  \bibinfo{journal}{Phys.Rev.Lett.} \textbf{\bibinfo{volume}{109}},
  \bibinfo{pages}{152005} (\bibinfo{year}{2012}{\natexlab{a}}),
  \eprint{1202.2843}.

\bibitem[{\citenamefont{Kogut and Soper}(1970)}]{Kogut:1969xa}
\bibinfo{author}{\bibfnamefont{J.~B.} \bibnamefont{Kogut}} \bibnamefont{and}
  \bibinfo{author}{\bibfnamefont{D.~E.} \bibnamefont{Soper}},
  \bibinfo{journal}{Phys.Rev.} \textbf{\bibinfo{volume}{D1}},
  \bibinfo{pages}{2901} (\bibinfo{year}{1970}).

\bibitem[{\citenamefont{Weinberg}(1995)}]{Weinberg:1995mt}
\bibinfo{author}{\bibfnamefont{S.}~\bibnamefont{Weinberg}}
  (\bibinfo{year}{1995}).

\bibitem[{\citenamefont{Brodsky et~al.}(1998)\citenamefont{Brodsky, Pauli, and
  Pinsky}}]{Brodsky:1997de}
\bibinfo{author}{\bibfnamefont{S.~J.} \bibnamefont{Brodsky}},
  \bibinfo{author}{\bibfnamefont{H.-C.} \bibnamefont{Pauli}}, \bibnamefont{and}
  \bibinfo{author}{\bibfnamefont{S.~S.} \bibnamefont{Pinsky}},
  \bibinfo{journal}{Phys.Rept.} \textbf{\bibinfo{volume}{301}},
  \bibinfo{pages}{299} (\bibinfo{year}{1998}), \eprint{hep-ph/9705477}.

\bibitem[{\citenamefont{Jaffe and Manohar}(1990)}]{Jaffe:1989jz}
\bibinfo{author}{\bibfnamefont{R.~L.} \bibnamefont{Jaffe}} \bibnamefont{and}
  \bibinfo{author}{\bibfnamefont{A.}~\bibnamefont{Manohar}},
  \bibinfo{journal}{Nucl. Phys.} \textbf{\bibinfo{volume}{B337}},
  \bibinfo{pages}{509} (\bibinfo{year}{1990}).

\bibitem[{\citenamefont{Lorce and Pasquini}(2011)}]{Lorce:2011kd}
\bibinfo{author}{\bibfnamefont{C.}~\bibnamefont{Lorce}} \bibnamefont{and}
  \bibinfo{author}{\bibfnamefont{B.}~\bibnamefont{Pasquini}},
  \bibinfo{journal}{Phys.Rev.} \textbf{\bibinfo{volume}{D84}},
  \bibinfo{pages}{014015} (\bibinfo{year}{2011}), \eprint{1106.0139}.

\bibitem[{\citenamefont{Lorce et~al.}(2012)\citenamefont{Lorce, Pasquini,
  Xiong, and Yuan}}]{Lorce:2011ni}
\bibinfo{author}{\bibfnamefont{C.}~\bibnamefont{Lorce}},
  \bibinfo{author}{\bibfnamefont{B.}~\bibnamefont{Pasquini}},
  \bibinfo{author}{\bibfnamefont{X.}~\bibnamefont{Xiong}}, \bibnamefont{and}
  \bibinfo{author}{\bibfnamefont{F.}~\bibnamefont{Yuan}},
  \bibinfo{journal}{Phys.Rev.} \textbf{\bibinfo{volume}{D85}},
  \bibinfo{pages}{114006} (\bibinfo{year}{2012}), \eprint{1111.4827}.

\bibitem[{\citenamefont{Lorce}(2012{\natexlab{a}})}]{Lorce:2012rr}
\bibinfo{author}{\bibfnamefont{C.}~\bibnamefont{Lorce}}
  (\bibinfo{year}{2012}{\natexlab{a}}), \eprint{1205.6483}.

\bibitem[{\citenamefont{Lorce}(2012{\natexlab{b}})}]{Lorce:2012ce}
\bibinfo{author}{\bibfnamefont{C.}~\bibnamefont{Lorce}}
  (\bibinfo{year}{2012}{\natexlab{b}}), \eprint{1210.2581}.

\bibitem[{\citenamefont{Hatta}(2012)}]{Hatta:2011ku}
\bibinfo{author}{\bibfnamefont{Y.}~\bibnamefont{Hatta}},
  \bibinfo{journal}{Phys.Lett.} \textbf{\bibinfo{volume}{B708}},
  \bibinfo{pages}{186} (\bibinfo{year}{2012}), \eprint{1111.3547}.

\bibitem[{\citenamefont{Ji et~al.}(2012{\natexlab{b}})\citenamefont{Ji, Xiong,
  and Yuan}}]{Ji:2012vj}
\bibinfo{author}{\bibfnamefont{X.}~\bibnamefont{Ji}},
  \bibinfo{author}{\bibfnamefont{X.}~\bibnamefont{Xiong}}, \bibnamefont{and}
  \bibinfo{author}{\bibfnamefont{F.}~\bibnamefont{Yuan}},
  \bibinfo{journal}{Phys.Lett.} \textbf{\bibinfo{volume}{B717}},
  \bibinfo{pages}{214} (\bibinfo{year}{2012}{\natexlab{b}}),
  \eprint{1209.3246}.

\bibitem[{\citenamefont{Ji et~al.}(2012{\natexlab{c}})\citenamefont{Ji, Xiong,
  and Yuan}}]{Ji:2012ba}
\bibinfo{author}{\bibfnamefont{X.}~\bibnamefont{Ji}},
  \bibinfo{author}{\bibfnamefont{X.}~\bibnamefont{Xiong}}, \bibnamefont{and}
  \bibinfo{author}{\bibfnamefont{F.}~\bibnamefont{Yuan}}
  (\bibinfo{year}{2012}{\natexlab{c}}), \eprint{1207.5221}.

\bibitem[{\citenamefont{Leader}(2012)}]{Leader:2011cr}
\bibinfo{author}{\bibfnamefont{E.}~\bibnamefont{Leader}},
  \bibinfo{journal}{Phys.Rev.} \textbf{\bibinfo{volume}{D85}},
  \bibinfo{pages}{051501} (\bibinfo{year}{2012}), \eprint{1109.1230}.

\end{thebibliography}
\end{document}